\documentclass[12pt,a4paper]{article}
\usepackage{amsmath}
\usepackage{amsfonts}
\usepackage{amssymb}
\usepackage{graphicx}
\usepackage{tikz}

\usepackage{caption}
\usepackage{subcaption}
\usepackage{sidecap}
\usepackage{color}
\usepackage{chngcntr}
\usepackage{cite}
\usepackage{authblk}
\counterwithout{figure}{section}

\usepackage{stackengine}
\stackMath
\usepackage{amssymb}
\usepackage{amsmath}
\usepackage{color}
\usepackage[colorlinks=true,allcolors=blue]{hyperref}

\title{Coherent states in magnetized anisotropic 2D Dirac materials} 
\date{ }
\author[1]{E D\'iaz-Bautista\footnote{ediaz@fis.cinvestav.mx}}
\author[1]{M Oliva-Leyva\footnote{moliva@fis.cinvestav.mx}}
\author[2]{Y Concha-S\'anchez\footnote{yconcha@umich.mx}}
\author[3]{A Raya\footnote{raya@ifm.umich.mx}}
\affil[1]{\small Physics Department, Cinvestav, P.O. Box. 14-740, 07000 Mexico City, Mexico}
\affil[2]{\small Facultad de Ingenier\'{\i}a Civil, Universidad Michoacana de San Nicol\'as de Hidalgo, Edificio C, Ciudad Universitaria. Francisco J. M\'ujica s/n. Col. Fel\'{\i}citas del R\'{\i}o. 58030 Morelia, Michoac\'an, M\'exico}
\affil[3]{\small Instituto de F\'{\i}sica y Matem\'aticas, Universidad Michoacana de San Nicol\'as de Hidalgo, Edificio C-3, Ciudad Universitaria. Francisco J. M\'ujica s/n. Col. Fel\'{\i}citas del R\'{\i}o. 58040 Morelia, Michoac\'an, M\'exico}
\affil[3]{\small Centro de Ciencias Exactas, Universidad del B\'io-B\'io, Avda. Andr\'es Bello 720, Casilla 447, 3800708, Chill\'an, Chile}
\begin{document}
	\maketitle

\begin{abstract}
In this work, we construct coherent states for electrons in anisotropic 2D Dirac materials immersed in a uniform magnetic field orthogonal to the sample. In order to describe the bidimensional effects on electron dynamics in a semiclassical approach, we adopt the symmetric gauge vector potential to describe the magnetic field. By solving a Dirac-like equation with an anisotropic Fermi velocity, we define two sets of generalized ladder operators that are generators of either the Heisenberg-Weyl or {\it su}(1,1) algebra and construct coherent states as eigenstates of the generalized annihilation operators with complex eigenvalues. In order to illustrate the anisotropy effects on these states, we obtain their probability density and mean energy value. Depending upon the anisotropy, expressed by the ratio between the Fermi velocities along the $x$- and $y$-axes, the shape of the probability density is modified on the $xy$-plane with respect to the isotropic case and according to the classical dynamics.
\end{abstract}


\section{Introduction}\label{sec1}

Material science has experienced a tremendous revolution after the first isolation of graphene samples~\cite{ngmzd04,ztsk05,cngpn09}, the first member of a larger class of materials nowadays dubbed generically as 2D Dirac materials, which include topological insulators \cite{hk10,qz11} and organic conductors \cite{kks09,kntsk14}. In general, these materials are characterized because at low energy, the behavior of its charge carriers is quite similar to that of ultra-relativistic fermions, given the linear nature of its dispersion relation.~As a consequence, these quasi-particles are better described by a Dirac-like equation, instead of the ordinary Schr\"{o}dinger equation with a typical parabolic dispersion relation. Such a description encodes the chiral and pseudo-relativistic behavior of the charge carriers that make it difficult to confine these quasi-particles through, for example, electrostatic barriers. Various proposals have been raised to address this problem, among them, to induce quasi-bounded states of massless Dirac fermions by the influence of magnetic fields. Another alternative to control confinement and transport of charge carriers in materials that are not intrinsically anisotropic, like graphene, comes from strain engineering. In this connection, {\em straintronics}~\cite{pcp09} has emerged as the field that explores how mechanical deformations of graphene flakes modify its electric properties~\cite{gbot17}. Because, in general, the strain tensor depends on the coordinates of the membrane, one particular effect arising from mechanical deformation is that an applied strain causes electrons to behave as if they were immersed in a fictitious magnetic field, a situation that has provided interesting theoretical~\cite{gbot17} and experimental~\cite{tppj09} results. In contrast, uniform uniaxial strain modifies the Fermi velocity in the low-energy regime without generating any pseudo-magnetic fields~\cite{og13,bccc15,ow17-1} but still induces a tensor character to the Fermi velocity in the material and consequently, the dispersion relation is modified from the ideal case and the low-energy regime corrects the equations of motion still in a tractable form.

Nowadays, an increasing interest in exploring the anisotropy effects to control other physical properties of 2D Dirac materials, {e.g.}, its rigidity, resistance and optical conductivity~\cite{ow17}, has arisen. In this way, the construction of coherent states in pristine graphene as carried out in Ref.~\cite{df17} can be extended to anisotropic Dirac fermion systems in order to give a semi-classical description of phenomena related to the combined effects of both magnetic fields and anisotropy~\cite{cdr19}. The latter allows to analyze a variety of properties of these materials~\cite{ks85,fk70,acg72,wh73,zfg90,aag00}.

Following the previous approach, the physical problem of a spinless particle moving in the $xy$-plane under the action of a uniform magnetic field $\vec{B}_0$ has been solved in the so-called symmetric gauge \cite{f28,p30,d31}
\begin{equation}\label{1}
\vec{A}=\frac12[\vec{B}_0\times\vec{r}]=\frac{B_0}{2}(-y,x,0),
\end{equation}

\begin{figure*}[!ht]
	\centering
	\includegraphics[width=0.7\linewidth]{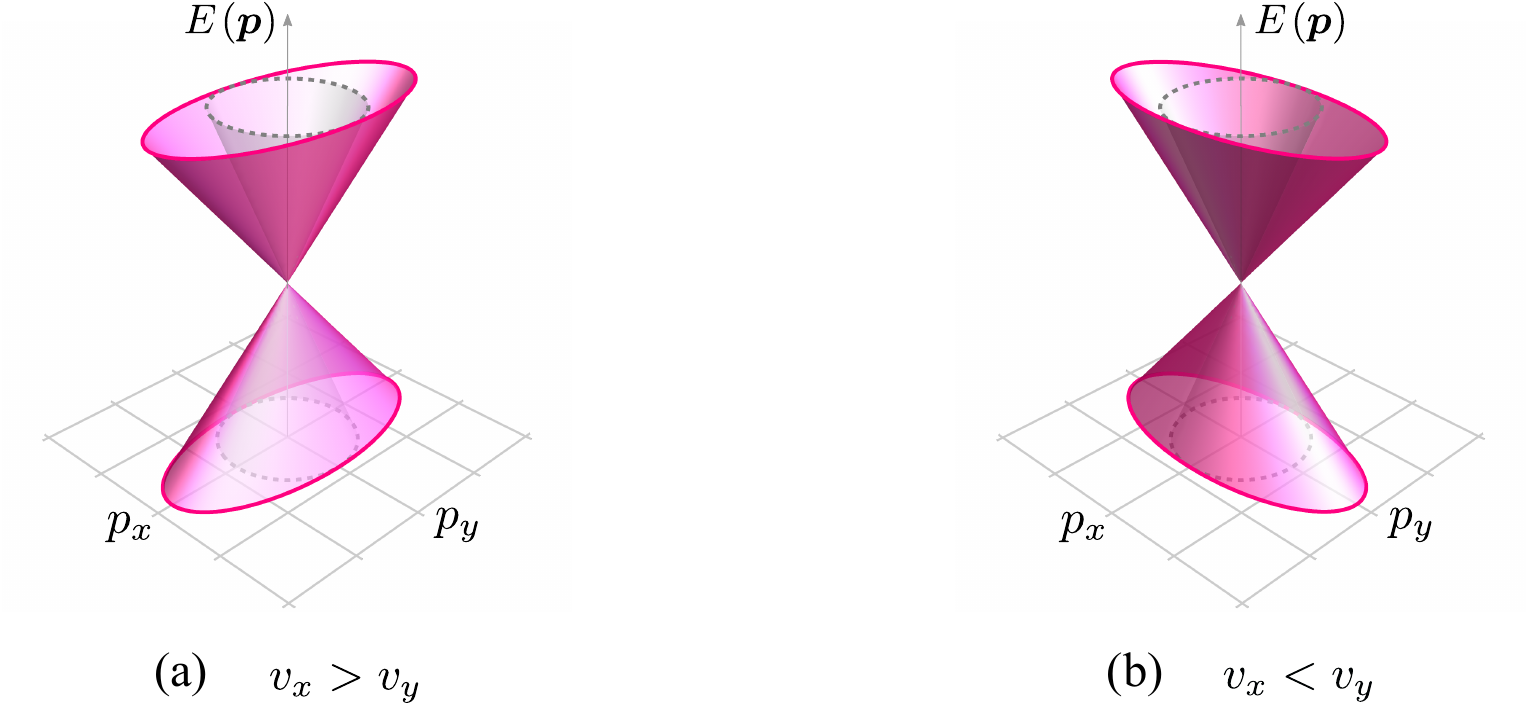}
	\caption{\label{fig:lattice}Dirac cones for an isotropic (dashed gray lines) and anisotropic case (solid pink contour). Dirac cone projections on the horizontal plane are ellipses whose semi-major axis is along either of the $p_x$-axis or of the $p_y$-axis.}
\end{figure*}

\noindent and in which has made possible to build coherent states~\cite{mm69} as bidimensional generalizations of the Glauber states~\cite{g63}, starting from the results obtained by Landau~\cite{landau30}. Thus, one can seek to generalize this formalism in the case of anisotropic 2D Dirac materials considering initially homogeneous perpendicular magnetic fields for describing the charge carrier dynamics as in Ref.~\cite{cdr19}, where the problem has been addressed in a Landau-like gauge with translational invariance and that conforms a first attempt for a semi-classical formulation in this kind of materials. However, if one would want to apply the coherent state formalism to investigate, for instance, some thermodynamical quantities of interest in magnetized systems, such description is limited due to the intrinsic bidimensional nature of the problem, so that it is necessary to implement the symmetric gauge~\cite{fk70,Rasetti1975,Cabrera1994,Gazeau00,Gazeau02,li18}. For instance, the gauge~(\ref{1}) has been employed to study some properties of an fermionic atoms trapped in an optical square lattice subjected to an external and classical non-Abelian gauge field~\cite{goldman09}. Therefore, as is indicated in~\cite{aremua15}, the coherent state basis on the symmetric gauge can be used to study Landau diamagnetism and de Haas-van Alphen oscillations, as well as in the calculation of the partition function~\cite{fk70} for a condensed matter system, as 2D Dirac materials.

Hence, we start form the time-independent anisotropic Dirac equation~\cite{kks09,og13,goldman09},
\begin{equation}
H\Psi=\left(v_{x}\sigma_x \pi_x +v_{y}\sigma_y\pi_y\right)\Psi=E\Psi,\label{2}
\end{equation}
where $v_{x/y}$ is the particle velocity in the $x/y-$direction and that is modified by the anisotropy --although the Fermi velocity has a tensorial character in a general sense, one can choose an appropriate coordinate system where the tensor is diagonal~\cite{ow17}--, $\sigma_{x/y}$ are the Pauli matrices, $\pi_{x/y}=\left(p_{x,y}+(e/c)A_{x/y}\right)$ is the canonical momentum with $\vec A$ given in Eq.~(\ref{1}) and $\Psi=\left(\begin{array}{c c}
\psi_1 & \psi_2
\end{array}\right)^{\rm T}$ are the two-component wavefunctions. This framework can be considered as a generalization of Refs.~\cite{df17,dnn19}. From here, we start the construction of coherent states, which we outline in the remaining of this article, organized as follows: In Sec.~\ref{sec2}, the anisotropic Dirac equation with a perpendicular constant magnetic field is solved in the symmetric gauge with an anisotropy dimensionless parameter, and its associated algebraic structure is discussed. In Sec.~\ref{sec3}, coherent states for graphene that preserve a rotational-like invariance are presented as eigenstates of matrix annihilation operators.~Also, the corresponding probability and mean energy are evaluated. 
Our final remarks are presented in Sec.~\ref{sec5}.

\section{Anisotropic Dirac Hamiltonian}\label{sec2}

By considering the symmetric gauge vector potential~(\ref{1}) in Eq.~(\ref{2}), Dirac Hamiltonian $H$ is rewritten as follows
\begin{equation}\label{3}
H=\sqrt{\omega_{\rm B}}\hbar\,v'_{\rm F}\left[\begin{array}{cc}
0 & -iA^{-} \\
iA^{+} & 0
\end{array}\right],
\end{equation}
after defining the ladder operators
\begin{equation}
A^{\pm}=\frac{\mp\,i}{\sqrt{\omega_{\rm B}}\hbar}\Bigg[\zeta^{1/2}\left(p_x-\frac{eB_0}{2c}y\right)\pm \frac{i}{\zeta^{1/2}}\left(p_y+\frac{eB_0}{2c}x\right)\Bigg],\label{4}
\end{equation}
that satisfy the commutation relation
\begin{equation}\label{5}
[A^{-},A^{+}]=\mathbf{1},
\end{equation}
where $\omega_{\rm B}=2eB_0/c\hbar$ is the cyclotron frequency, $v'_{\rm F}=\sqrt{v_{x}v_{y}}$ and $\zeta=v_{x}/v_{y}$ depends on the anisotropy direction (see Fig.~\ref{fig:lattice}).


The eigenvalue equation~(\ref{3}) gives place to two coupled equations that
can be decoupled to obtain the following equations for each pseudo-spinor component,
\begin{subequations}\label{7}
	\begin{align}
	\mathcal{H}^-\psi_1(x,y)&=A^{-}A^{+}\psi_1(x,y)=\mathcal{E}\psi_1(x,y), \label{8a} \\
	\mathcal{H}^+\psi_2(x,y)&=A^{+}A^{-}\psi_2(x,y)=\mathcal{E}\psi_2(x,y), \label{8b}
	\end{align}
\end{subequations}
with $\mathcal{E}\equiv(E/\hbar\,v'_{\rm F}\sqrt{\omega_{\rm B}})^2$. Thus, we have two Schr\"{o}dinger equations, each corresponding to a harmonic oscillator, but whose eigenvalues are related as
\begin{equation}
\mathcal{E}_{1,n-1}=\mathcal{E}_{2,n}=n, \quad n\geq 1, \quad \mathcal{E}_{2,0}=0,
\end{equation}
such that the energy spectrum turns out to be
\begin{equation}\label{11}
E_n=\pm\hbar v'_{\rm F}\sqrt{n\,\omega_{\rm B}}, \quad n=0,1,2,\dots,
\end{equation}
where the positive (negative) energy corresponds to electrons in the conduction (valence) band. As a part of our discussion, we just consider electrons in the conduction band.

Proceeding as in Ref.~\cite{dknn17}, the normalized eigenfunctions of the Hamiltonian $\mathcal{H}^+$ turn out to be \cite{f28}
\begin{align}\label{34}
\nonumber\psi_{m,n}(\rho,\theta)&=(-1)^{\min(m,n)}\sqrt{\frac{\omega_{\rm B}}{4\pi}\frac{\min(m,n)!}{\max(m,n)!}}\left(\frac{\sqrt{\omega_{\rm B}}}{2}\rho\right)^{\vert n-m\vert}\exp\left(-\frac{\omega_{\rm B}}{8}\rho^2+i(n-m)\theta\right)\times\\
&\quad\times L_{\min(m,n)}^{\vert n-m\vert}\left(\frac{\omega_{\rm B}}{4}\rho^2\right),
\end{align}
with $n,m=0,1,2,\dots$ and $L_a^b(x)$ representing the associated Laguerre polynomials. Notice that the normalized eigenfunctions of the Hamiltonian $\mathcal{H}^-$ are obtained simply as $\psi_{m,n-1}=A^{-}\psi_{m,n}/\sqrt{n}$.

As we can see, the eigenstates of the Hamiltonians $\mathcal{H}^\pm$ are labeled by two positive, integer numbers $m,n$, that correspond to the eigenvalues of two number operators $M$ and $N$, respectively (see Appendix~\ref{appA}):
\begin{equation}\label{28}
\psi_1(\xi,\theta)\equiv\psi_{m,n-1}(\xi,\theta), \quad \psi_2(\xi,\theta)\equiv\psi_{m,n}(\xi,\theta).
\end{equation}
From Eq.~(\ref{21c}), one concludes that the states $\psi_{m,n}$ are also eigenstates of the angular momentum-like operator $L_z=N-M$ with eigenvalue $m_z=n-m$. 

Finally, by defining the $z$-component of the total angular momentum operator as $\mathbb{J}_z=L_z\otimes\mathbb{I}+\sigma_z/2$, we have that
\begin{align}\label{18}
\mathbb{J}_z\Psi_{m,n}(x,y)&=\left(m_z-\frac12\right)\Psi_{m,n}(x,y)=j\,\Psi_{m,n}(x,y),
\end{align}
{\it i.e.}, the states $\Psi_{m,n}(x,y)$ are also eigenstates of $\mathbb{J}_z$ with rational eigenvalue $j\equiv m_z-1/2$.

In Appendix~\ref{class}, we establish a classification of the states $\Psi_{m,n}(x,y)$ according to the sign of $m_z$ as well as the Hilbert space spanned and the completeness relation.

\section{Annihilation operators}\label{sec3}
Let us define two independent annihilation operators $\mathbb{A}^{-}$ and $\mathbb{B}^{-}$ as
\begin{subequations}\label{44.1}
	\begin{align}
	\mathbb{A}^{-}&=\left[\begin{array}{c c}
	\cos(\delta)\frac{\sqrt{N+2}}{\sqrt{N+1}}A^{-} & \sin(\delta)\frac{1}{\sqrt{N+1}}(A^{-})^2 \\
	-\sin(\delta)\sqrt{N+1} & \cos(\delta)A^{-}
	\end{array}\right], \label{19a}\\ 
	\mathbb{B}^{-}&=\left[\begin{array}{c c}
	\cos(\eta) B^{-} & \sin(\eta)\frac{B^{-}}{\sqrt{N+1}}A^{-} \\
	-\sin(\eta)A^{+}\frac{B^{-}}{\sqrt{N+1}} & \cos(\eta)B^{-}
	\end{array}\right], \label{19b}
	\end{align}
\end{subequations}
such that $\delta,\,\eta\in[0,2\pi]$ and for $m,n\in\mathbb{Z}^+\cup\{0\}$:
\begin{subequations}
	\begin{align}
	\mathbb{A}^{-}\Psi_{m,n}&=\frac{\exp(i\delta)}{\sqrt{2^{\delta_{1n}}}}\sqrt{n}\Psi_{m,n-1}, \\
	\mathbb{B}^{-}\Psi_{m,n}&=\sqrt{m}\,\omega_n\Psi_{m-1,n},
	\end{align}
\end{subequations}
with $\omega_n\equiv\cos(\eta)+i\sin(\eta)(1-\delta_{0n})$.

\begin{figure}[!ht]
	\centering
	\includegraphics[width=.45\columnwidth]{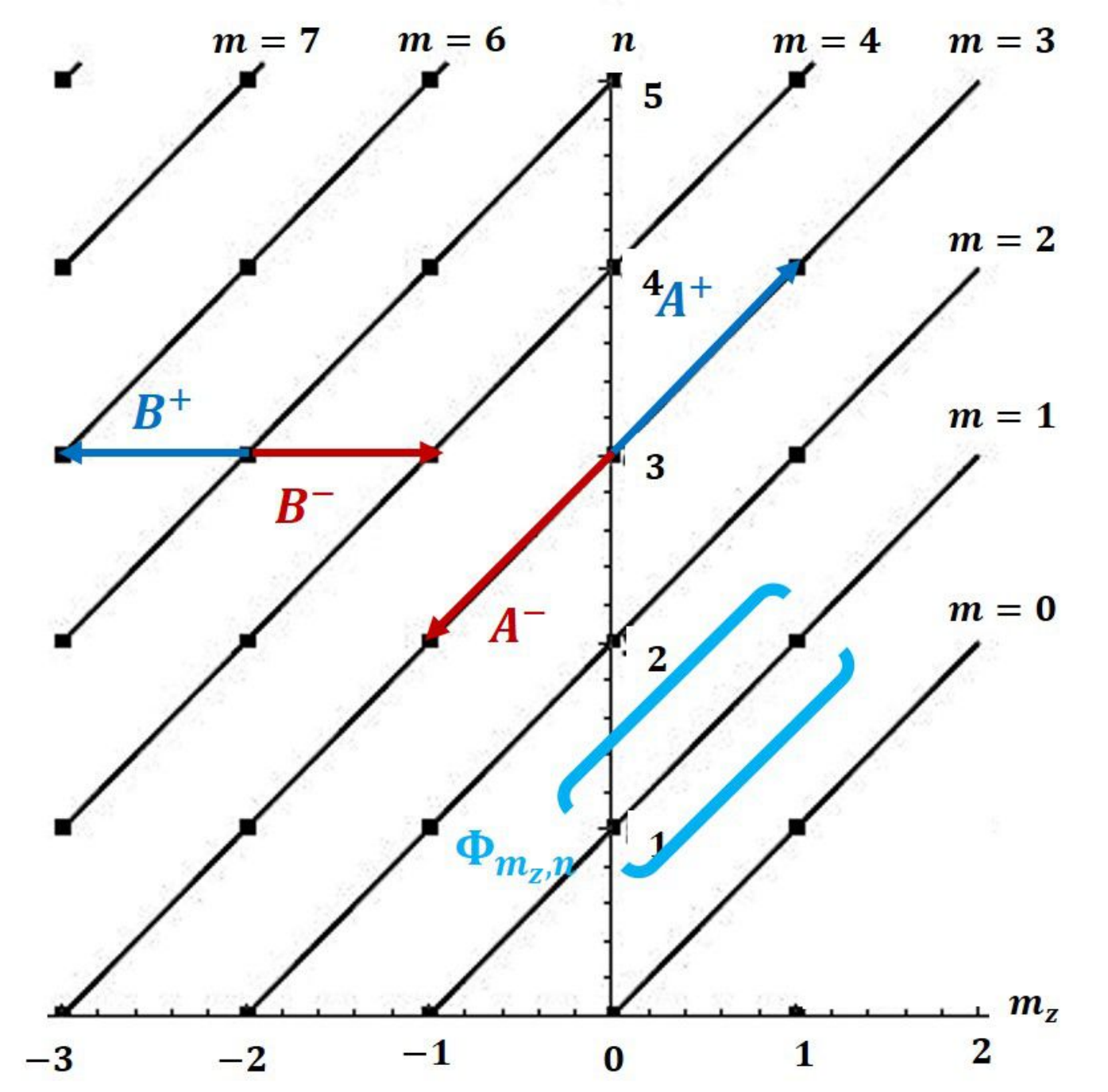}
	\caption{\label{fig:diagramLN}Space of scalar states $\phi_{m_z,n}$ is represented univocally by coordinates $(m_z,n)$. Inclined lines connect states with the value $m=n-m_z$. The plane is divided in two sectors according to $m_z\geq0$ (right sector) or $m_z<0$ (left sector).}
\end{figure}

Additionally, let us consider the following operator
\begin{equation}
\label{24}
\mathbb{K}^{-}=\mathbb{A}^{-}\mathbb{B}^{-}\equiv\left[\begin{array}{c c}
\cos(\gamma)\frac{\sqrt{N+2}}{\sqrt{N+1}}A^{-}B^{-} & \sin(\gamma)\frac{1}{\sqrt{N+1}}(A^{-})^2B^{-} \\
-\sin(\gamma)\sqrt{N+1}B^{-} & \cos(\gamma)A^{-}B^{-}
\end{array}\right],
\end{equation}
where $\gamma=\delta+\eta\in[0,2\pi]$ --due to the periodicity of sine and cosine functions-- such that for $m_z=0,\pm 1,\pm 2,\dots$, it verifies
\begin{equation}\label{25}
\mathbb{K}^{-}\Phi_{m_z,n}=\frac{\exp(i\gamma)}{\sqrt{2^{\delta_{1n}}}}\sqrt{n(n-m_z)}\Phi_{m_z,n-1},  
\end{equation}
for $n=0,1,2,\dots,$ with (see Fig.~\ref{fig:diagramLN})
\begin{equation}
\Phi_{m_z,n}(x,y)=\frac{1}{\sqrt{2^{(1-\delta_{0n})}}}\left(\begin{array}{c}
(1-\delta_{0n})\phi_{m_z-1,n-1}(x,y) \\
i\phi_{m_z,n}(x,y)
\end{array}\right).
\end{equation}

It is worth to remark that depending on the values of $\delta,\,\eta$ and $\gamma$, we can work with either diagonal or non-diagonal matrix operators.

On the other hand, bidimensional coherent states $\Psi_{\alpha,\beta}(x,y)\equiv\langle x,y\vert\alpha,\beta\rangle$ are defined as common eigenstates of the generalized annihilation operators $\mathbb{A}^{-}$ and $\mathbb{B}^{-}$ \cite{mm69,d17,dknn17}:
\begin{subequations}\label{19}
	\begin{align}
	\mathbb{A}^{-}\Psi_{\alpha,\beta}(x,y)&=\alpha\Psi_{\alpha,\beta}(x,y), \quad \alpha\in\mathbb{C}, \\ \mathbb{B}^{-}\Psi_{\alpha,\beta}(x,y)&=\beta\Psi_{\alpha,\beta}(x,y), \quad \beta\in\mathbb{C}.
	\end{align}
\end{subequations}

Similarly, it is possible to construct another kind of coherent states, $\Phi_{\tau}^{m_z}(x,y)\equiv\langle x,y\vert\tau,m_z=n-m\rangle$ which are also eigenstates of the operator $\mathbb{J}_z$, {\it i.e.},
\begin{subequations}\label{27}
	\begin{align}
	\mathbb{K}^{-}\Phi_{\tau}^{m_z}(x,y)&=\tau\Phi_{\tau}^{m_z}(x,y), \quad \tau\in\mathbb{C}, \label{27a} \\ \mathbb{J}_z\Phi_{\tau}^{m_z}(x,y)&=j\,\Phi_{\tau}^{m_z}(x,y), \quad j=\pm\frac12,\pm\frac32,\pm\frac52,\dots. \label{27b}
	\end{align}
\end{subequations}

A brief discussion regarding the algebras associated with these annihilation operators can be found in Appendix~\ref{algebra}. In the forthcoming sections, we build bidimensional and {\it su}(1,1) coherent states \cite{ng03,dhm12,dm13} in graphene for some particular values of the parameters $\delta,\,\eta$ and $\gamma$. These states belong to a kind of coherent states known in literature as vector coherent states (VCS)~\cite{aag00,ali03,ali04,ali05,bagarello09} since they are defined over matrix domains and, generally, they are multicomponent coherent states, $\vert q,k\rangle$, where $q$ ranges through some continuous parameter space and $k$ is a finite discrete index.

\subsection{Bidimensional coherent states (2D-CS)}\label{sec3.3}
In general, bidimensional states are a linear combination of all the stationary states $\Psi_{m,n}(x,y)$ \cite{d17}:
\begin{equation}
\Psi_{\alpha,\beta}(x,y)=\mathcal{N}\sum_{n=0}^{\infty}\sum_{m=0}^{\infty}c_nd_m\Psi_{m,n}(x,y)
=\mathcal{N}\sum_{n=0}^{\infty}c_n\Psi_{\beta}^n(x,y)=\mathcal{N}\sum_{m=0}^{\infty}d_m\Psi_{\alpha}^m(x,y), \label{93}
\end{equation}
where $\mathcal{N}$ is a normalization constant and the states $\Psi_{\alpha}^m(x,y)$, $\Psi_{\beta}^n(x,y)$ are the respective eigenstates of the operators $\mathbb{A}^{-}$ and $\mathbb{B}^{-}$ {\it i.e.},
\begin{subequations}
	\begin{align}
\mathbb{A}^{-}\Psi_{\alpha}^m(x,y)&=\alpha\Psi_{\alpha}^m(x,y), \\
 \mathbb{B}^{-}\Psi_{\beta}^n(x,y)&=\beta\Psi_{\beta}^n(x,y).\label{29}
\end{align}
\end{subequations}
Therefore, Eq.~(\ref{93}) provides a way to obtain bidimensional coherent states building first either the states $\Psi_{\alpha}^m(x,y)$ or $\Psi_{\beta}^n(x,y)$ and then gathering them together properly (see Appendix~\ref{2DCS}).

Hence, we obtain the explicit expression for the bidimensional coherent states 
\begin{equation}\label{38}
\Psi_{\alpha,\beta}(x,y)=\frac{\exp\left(\left[\beta-\frac{z}{2}\right]z^\ast-\frac{\vert\beta\vert^2}{2}\right)}{\sqrt{\pi(2\exp(\vert\tilde{\alpha}\vert^2)-1)}}\sum_{n=0}^{\infty}\frac{\tilde{\alpha}^n}{n!}\left(\begin{array}{c}
\sqrt{n}(z-\beta)^{n-1}\\
i(z-\beta)^n
\end{array}\right),
\end{equation}
where $z=\frac{\sqrt{\omega_{\rm B}}}{2}\left(\zeta^{-1/2}x+i\zeta^{1/2}y\right)$ and $\tilde{\alpha}=\alpha\exp\left(-i\delta\right)$. It is straightforward to verify that these quantum states satisfy the eigenvalue equations in~(\ref{19}). Note that the $\delta$-parameter introduces a phase factor to the eigenvalue $\alpha$.

\begin{figure*}[t!]
	\centering
	\includegraphics[width=0.7\linewidth]{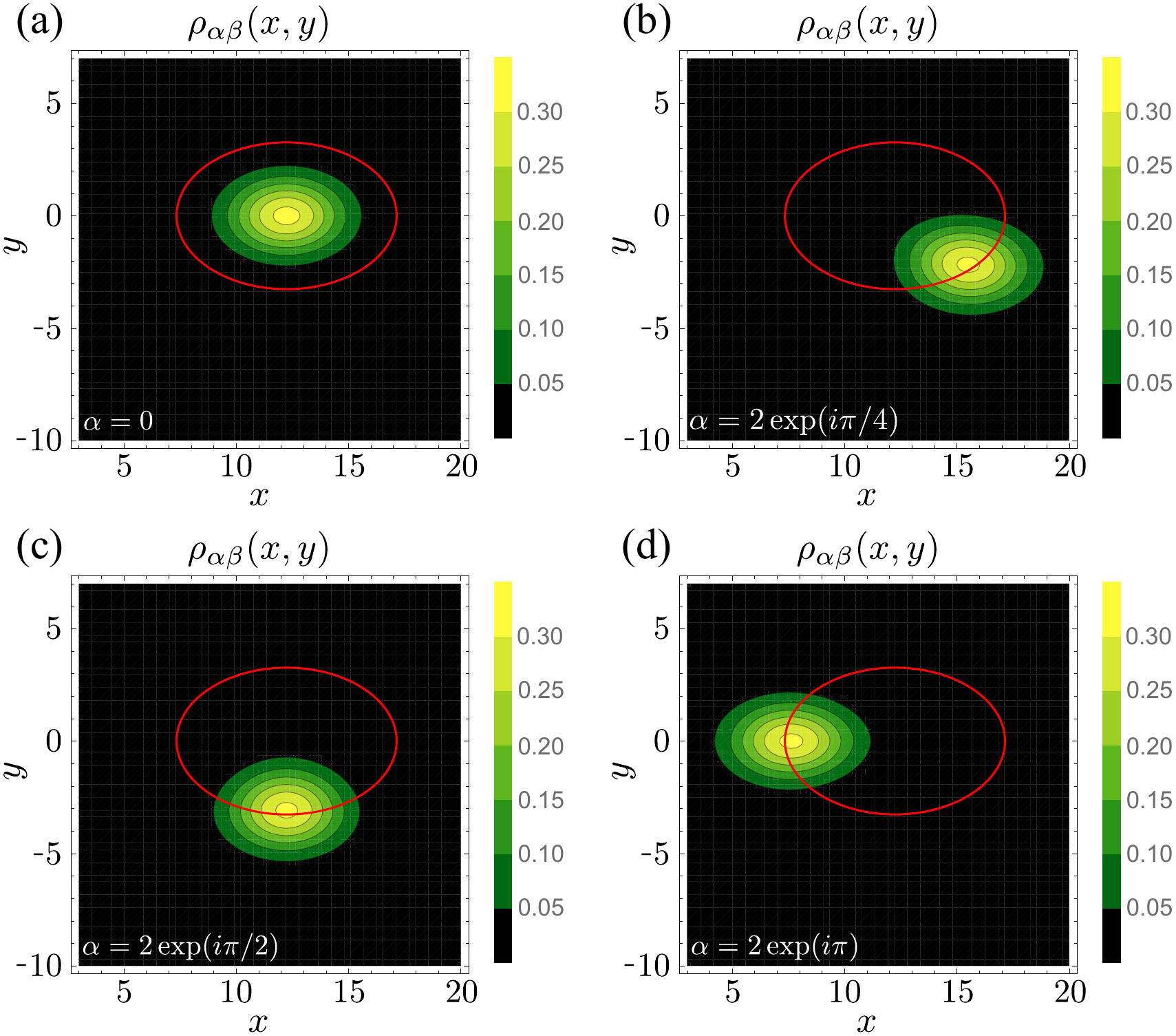}	
	\caption{\label{fig:rho_alphabeta1}For the 2D-CS $\Psi_{\alpha,\beta}(x,y)$, the probability density $\rho_{\alpha,\beta}(x,y)$ is shown for some values of $\alpha$. In all the cases $\beta=5$, $B_0=\zeta=1/2$ and $\omega_{\rm B}=1$. Red curves on the $xy$-plane describe the classical trajectory of a charge carrier in a magnetic field: the coordinates of the ellipse center are determined by $\beta$ while $\alpha$ gives the coordinates in which the maximum probability amplitude can be found respect to that point.}
\end{figure*}

\begin{figure*}[t!]
	\centering
	\includegraphics[width=0.7\linewidth]{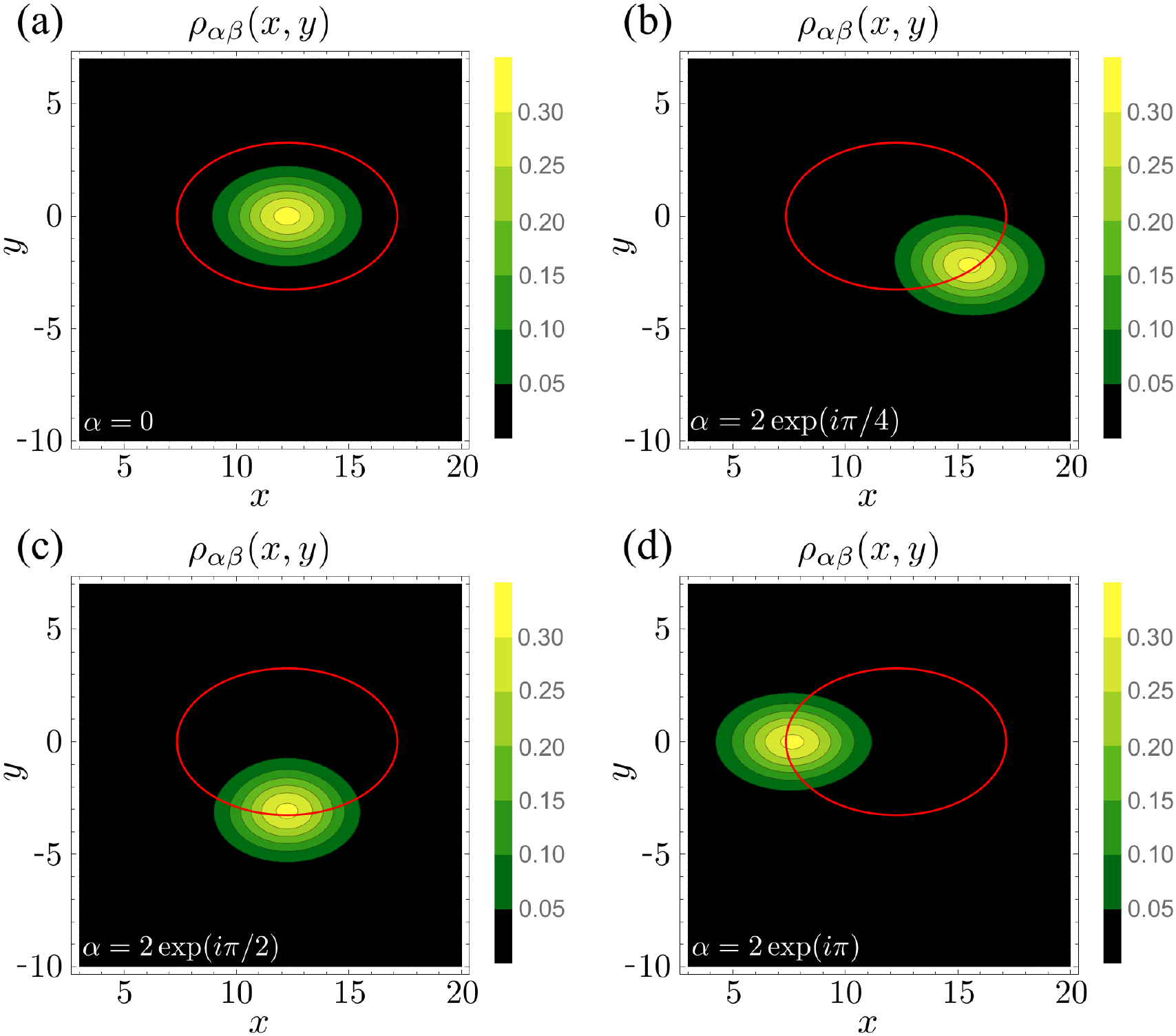}		
	\caption{\label{fig:rho_alphabeta}For the 2D-CS $\Psi_{\alpha,\beta}(x,y)$, the probability density $\rho_{\alpha,\beta}(x,y)$ is shown for some values of $\alpha$. In all the cases $\beta=5$, $B_0=1/2$, $\zeta=3/2$ and $\omega_{\rm B}=1$. Red curves on the $xy$-plane describe the classical trajectory of a charge carrier in a magnetic field: the coordinates of the ellipse center are determined by $\beta$ while $\alpha$ gives the coordinates in which the maximum probability amplitude can be found respect to that point.}
\end{figure*}

\begin{figure}[t!]
	\centering
	\includegraphics[width=0.45\linewidth]{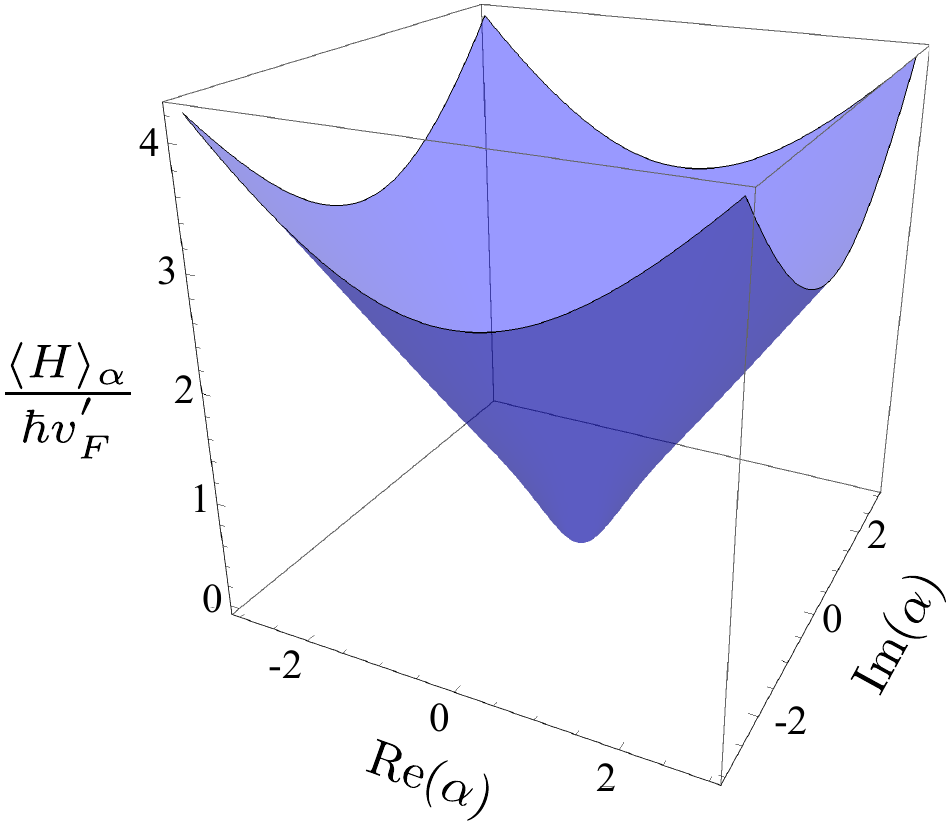}
	\caption{\label{fig:H_alpha}Mean energy $\langle H\rangle_{\alpha}/\hbar v'_{\rm F}$ as a function of $\alpha$ with $B_0=1/2$ and $\omega_{\rm B}=1$.}
\end{figure}

Finally, setting $\delta=0$, the corresponding probability density $\rho_{\alpha,\beta}(x,y)$ and mean energy value $\langle H\rangle_{\alpha}$ are, respectively (see Figs. \ref{fig:rho_alphabeta1}-\ref{fig:H_alpha}):
\footnotesize
	\begin{subequations}\label{94}
		\begin{align}
		\nonumber	\rho_{\alpha,\beta}(x,y)&=\Psi_{\alpha,\beta}^{\dagger}(x,y)\Psi_{\alpha,\beta}(x,y)\\
		&=\frac{\exp\left(-\vert z-\beta\vert^2\right)}{\pi(2\exp(\vert\alpha\vert^2)-1)}\Bigg[1+\left\vert\sum_{n=1}^{\infty}\frac{\left[\alpha(z-\beta)\right]^n}{n!}\right\vert^2+\left\vert\sum_{n=1}^{\infty}\frac{\left[\alpha(z-\beta)\right]^n\sqrt{n}}{n!(z-\beta)}\right\vert^2+2\Re\left[\sum_{n=1}^{\infty}\frac{\left[\alpha(z-\beta)\right]^n}{n!}\right]\Bigg], \label{94a} \\
		\langle H\rangle_\alpha&=\frac{2\sqrt{\omega_{\rm B}}\hbar v'_{\rm F}}{2\exp\left(\vert\alpha\vert^2\right)-1}\sum_{n=0}^{\infty}\frac{\vert\alpha\vert^{2n}}{n!}\sqrt{n}. \label{94c}
		\end{align}
	\end{subequations}
\normalsize

Figures \ref{fig:rho_alphabeta1} and \ref{fig:rho_alphabeta} show that the complex parameters $\alpha$ and $\beta$ determine jointly where the maximum probability density is on the $xy$-plane. Moreover, the probability density $\rho_{\alpha,\beta}(x,y)$ exhibits a stable shape regardless of the value of $\alpha$ and $\beta$, so that the behavior of $\rho_{\alpha,\beta}(x,y)$ function resembles the squeezed coherent states (SCS) in phase space representation, in which the product of the variances of the $x$-position and $p_{x}$-momentum operators minimize the Heisenberg uncertainty relation $(\Delta x)(\Delta p_{x})=1/4$ with $(\Delta x)^{2}=\exp\left(-2r\right)/4$ and $(\Delta p_{x})^2=\exp\left(2r\right)/4$, where $r=\vert\lambda\vert$ is called the squeezed parameter and $\lambda\in\mathbb{C}$. In our discussion, the quantity $\tau=\frac12\ln\left(\zeta\right)$ can be considered as an squeezed-like parameter for the $z$-variable.

Also, as we can see, the eigenvalues $\alpha$ and $\beta=\vert\beta\vert\exp(i\varphi)$ are related with the geometric parameters of the classical elliptic trajectory of a charged particle in a magnetic field on the $xy$-plane (see again Figs.~\ref{fig:rho_alphabeta1} and \ref{fig:rho_alphabeta}):
\begin{equation}
\frac{\left(x-x_0\right)^2}{4\zeta\vert\alpha\vert^2}+\frac{\left(y-y_0\right)^2}{4\zeta^{-1}\vert\alpha\vert^2}=1,
\end{equation}
where the points
\begin{equation}
(x_0,y_0)=\left(\frac{2\zeta^{1/2}\vert\beta\vert}{\sqrt{\omega_{\rm B}}}\cos(\varphi),\frac{2\zeta^{-1/2}\vert\beta\vert}{\sqrt{\omega_{\rm B}}}\sin(\varphi)\right),
\end{equation}
determine the coordinates of the center of the curve respect to the origin $(0,0)$, while the eccentricity $\varepsilon$ is given by
\begin{subequations}\label{eccentricity}
	\begin{align}
\varepsilon_{x}&=\sqrt{1-\zeta^{2}}, \hspace*{0.22cm}\quad {\rm for }\,\,v_{x}<v_{y},\\
\varepsilon_{y}&=\sqrt{1-\zeta^{-2}}, \quad {\rm for }\,\,v_{x}>v_{y}.
\end{align} 
\end{subequations}
Therefore, when the anisotropy is directed along the $x$-direction, the peak of maximum probability is found in elliptical curve whose semi-major axis is parallel to $y$-axis (Fig.~\ref{fig:rho_alphabeta1}), while if the anisotropy is along the $y$-direction, the semi-major axis is parallel to $x$-axis (Fig.~\ref{fig:rho_alphabeta}). Moreover, for the isotropic case we have that $\zeta=1$ and $\varepsilon_{x}=\varepsilon_{y}=0$, {\it i.e.}, we obtain a probability density $\rho_{\alpha,\beta}(x,y)$ that has a Gaussian-like shape and whose maximum value is located in a circular curve.

%

\subsection{{\it su}(1,1) coherent states ({\it su}(1,1)-CS)}\label{sec4}
Eq.~(\ref{25}) implies the states that are annihilated by the operator $\mathbb{K}^{-}$ are those  with the minimum energy values $E_0=0$ (for $m_z\leq0$) and $E_{m_z}=\hbar v'_{\rm F}\sqrt{m_z\omega}$ (for $m_z\geq0$) (see Fig.~\ref{fig:diagramLN}). Thus, for a fixed value of $m_z$, any state can be expressed as a linear combination of states labeled as $\Psi_{m_z,n}(x,y)$:
\begin{equation}\label{42}
\Phi^{m_z}(x,y)=\sum_{n=\max(0,m_z)}^{\infty}c_{m_z,n}\Phi_{m_z,n}(x,y),
\end{equation}
$m_z=0,\pm1,\pm2,\dots$, such that, under Eq.~(\ref{18}),
\begin{equation}
\mathbb{J}_z\Phi^{m_z}(x,y)=j\,\Phi^{m_z}(x,y), \quad j=m_z-\frac{1}{2}.
\end{equation}
This means that the coherent states $\Phi_{\tau}^{m_z}(x,y)$  have a defined angular momentum in $z$-direction (positive or negative sign) according to the sector to which the states $\Psi_{m_z,n}(x,y)$ that contribute to the superposition in Eq.~(\ref{42}) belong (right or left sector in Fig.~\ref{fig:diagramLN}).

As previous section, the states $\Phi_{\tau}^{m_z}(x,y)$ are obtained for the particular values $\eta=0$ and $\gamma=\delta$ in Eq.~(\ref{24}).

For each sector in Fig.~\ref{fig:diagramLN}, the {\it su}(1,1)-CS turn out to be, respectively:
\begin{subequations}
	\begin{align}
\Phi_{\tau}^{m_z\geq0}(x,y)&=\frac{1}{\sqrt{_0F_1\left(;m_z+1;\vert\tilde{\tau}\vert^2\right)}}\sum_{n=m_z}^{\infty}\frac{\sqrt{m_z!}\,\tilde{\tau}^{n-m_z}}{\sqrt{n!(n-m_z)!}}\Phi_{m_z,n}(x,y), \label{105} \\
\nonumber\Phi_{\tau}^{m_z<0}(x,y)&=\frac{1}{\sqrt{2 \, _0F_1\left(;-m_z+1;\vert \tilde{\tau}\vert^2\right)-1}}\times \\
&\quad\times\left[\Phi_{m_z,0}(x,y)+\sum_{n=1}^{\infty}\frac{\sqrt{2\,(-m_z)!}\tilde{\tau}^n}{\sqrt{n!(n-m_z)!}}\Phi_{m_z,n}(x,y)\right], \label{118}
\end{align}
\end{subequations}
where $\tilde{\tau}=\tau\exp(-i\delta)$ and $\,_pF_q$ is a generalized hypergeometric function defined by
\begin{equation}\label{106}
\,_pF_q(a_1,\dots,a_p,b_1,\dots,b_q;x)=\sum_{n=0}^{\infty}\frac{(a_1)_n\cdots(a_p)_n}{(b_1)_n\cdots(b_q)_n}\frac{x^n}{n!},
\end{equation}
where $(a)_k$ is the Pochhammer symbol,
\begin{equation}
(a)_k=\frac{\Gamma(a+k)}{\Gamma(a)}.
\end{equation}



\begin{figure*}[t!]
	\centering
	\includegraphics[width=0.9\linewidth]{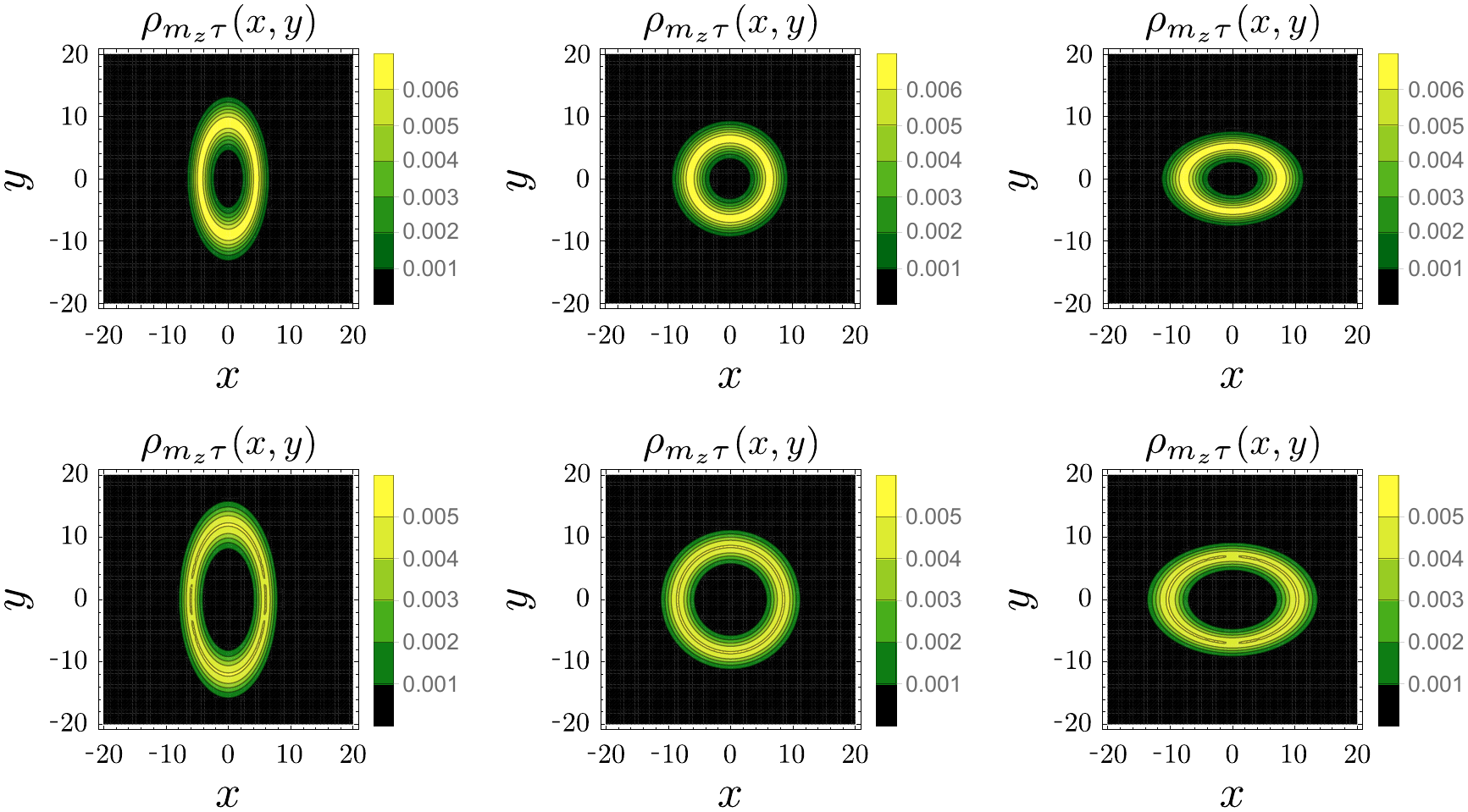}
	\caption{\label{fig:rho1P_z}Probability density $\rho_{m_z,\tau}(x,y)$ for the coherent states in Eq.~(\ref{105}) with $\vert j\vert=1/2$ for different values of the $\zeta$-parameter and $\vert\tau\vert$-eigenvalue: (vertical) $\zeta=1/2$, $1$, $3/2$, and (horizontal) $\vert\tau\vert=3$, $5$. In all these cases $B_0=1/2$ and $\omega_{\rm B}=1$.}
\end{figure*}

\begin{figure*}[t!]
	\centering
	\includegraphics[width=0.9\linewidth]{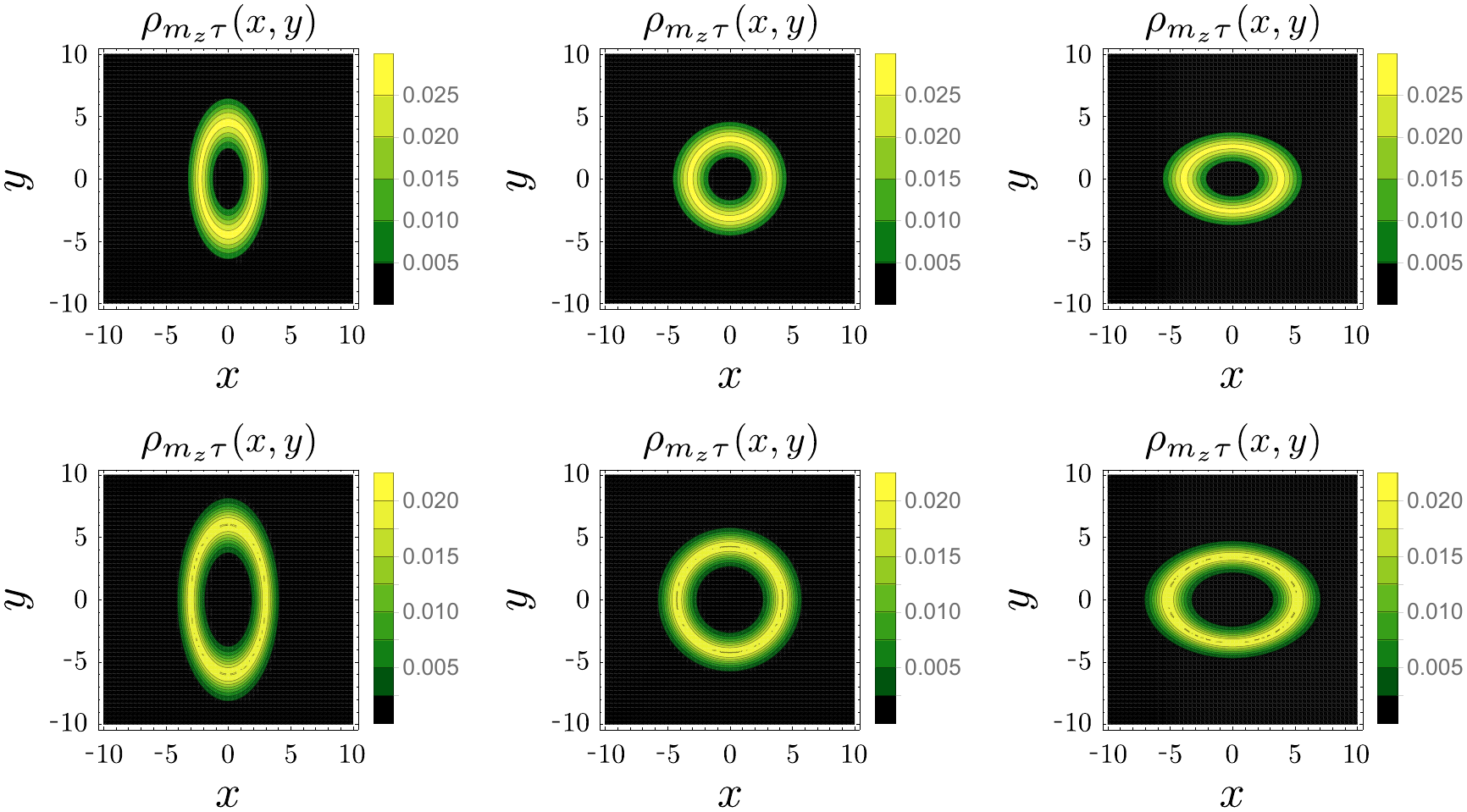}
	\caption{\label{fig:rho1N_z}Probability density $\rho_{m_z,\tau}(x,y)$ for the coherent states in Eq.~(\ref{118}) with $\vert j\vert=3/2$ for different values of the $\zeta$-parameter and $\vert\tau\vert$-eigenvalue: (vertical) $\zeta=1/2$, $1$, $3/2$, and (horizontal) $\vert\tau\vert=3$, $5$. In all these cases $B_0=2$ and $\omega_{\rm B}=2$.}
\end{figure*}

\begin{figure*}[t!]
	\centering
	\includegraphics[width=0.9\linewidth]{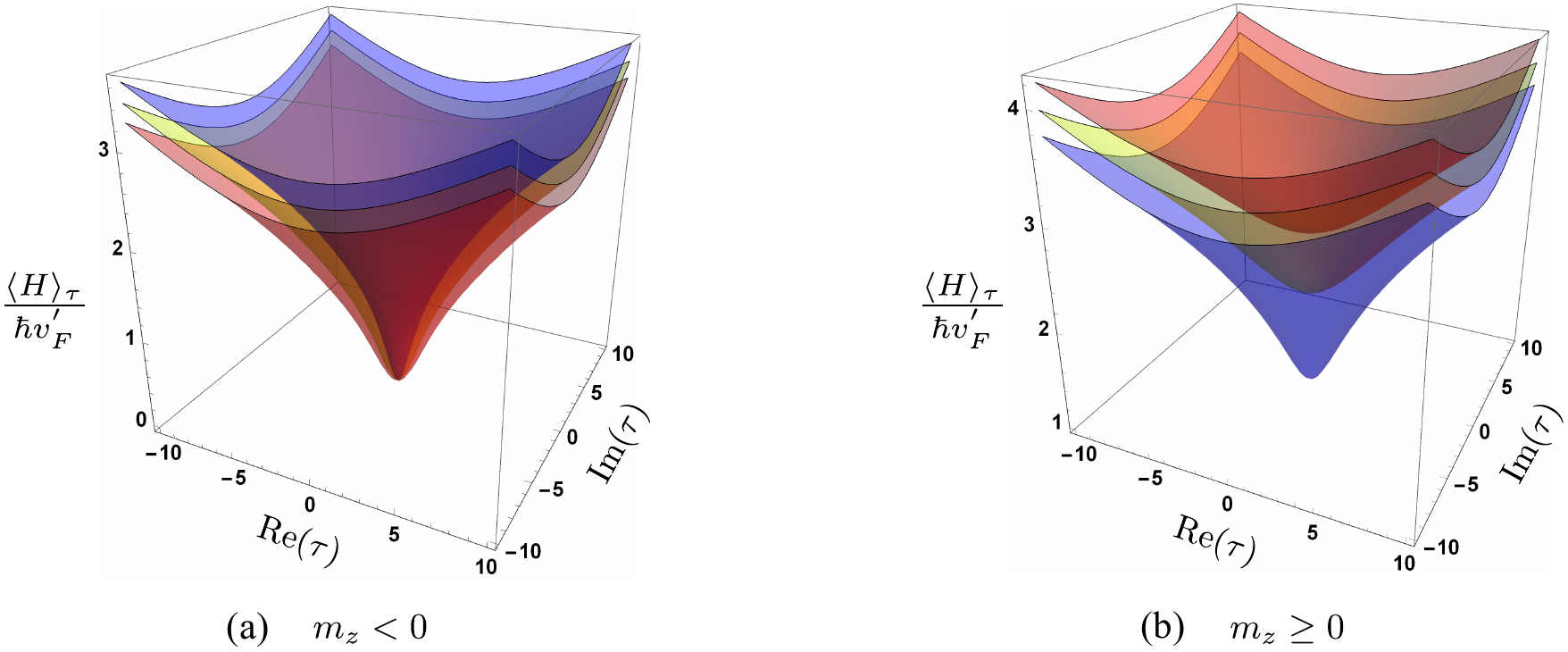}\\
	\caption{\label{fig:H_z}Mean energy value $\langle H\rangle_\tau/\hbar v'_{\rm F}$ as a function of $\tau$ for the {\it su}(1,1)-CS $\Phi_{\tau}^{m_z}(x,y)$ with $\vert m_z\vert=1$ (blue), $\vert m_z\vert=4$ (yellow) and $\vert m_z\vert=7$ (red). In all the cases $B_0=1/2$ and $\omega_{\rm B}=1$.}
\end{figure*}

Setting again $\delta=0$, the corresponding probability and mean energy value are given by (see Figs.~\ref{fig:rho1P_z}-\ref{fig:H_z}):
\small
\begin{subequations}\label{107}
	
		\begin{align}	
		\nonumber\rho_{m_z,\tau}(x,y)&=\rho_{m_z,\tau}(\xi)=\Phi_{\tau}^{m_z\dagger}(x,y)\Phi_\tau^{m_z}(x,y)=\frac{\omega_{\rm B}\,\vert z\vert^{2m_z}\exp\left(-\xi^2\right)}{8\pi\left[\, _0F_1\left(;m_z+1;\vert\tau\vert^2\right)\right]}\times \\
		&\quad\times\left[\left\vert\sum_{n=m_z}^{\infty}\frac{\sqrt{m_z!}(-\tau)^{n-m_z}}{n!}L_{n-m_z}^{m_z}(\xi^2)\right\vert^2+\left\vert\sum_{n=m_z}^{\infty}\frac{\sqrt{m_z!}(-\tau)^{n-m_z}z^{-1}}{\sqrt{n}(n-1)!}L_{n-m_z}^{m_z-1}(\xi^2)\right\vert^2\right],\label{119a} \\
		\langle H\rangle_\tau&=\frac{\sqrt{\omega_{\rm B}}\hbar v'_{\rm F}}{\, _0F_1\left(;m_z+1;\vert\tau\vert^2\right)}\sum_{n=m_z}^{\infty}\frac{m_z!\,\vert\tau\vert^{2n-2m_z}}{n!(n-m_z)!}\sqrt{n}, \quad {\rm for}\,m_z\geq0,\label{119c} \\
		%
		%
		%
		\nonumber\rho_{m_z,\tau}(x,y)&=\rho_{m_z,\tau}(\xi)=\Phi_{\tau}^{m_z\dagger}(x,y)\Psi_\tau^{m_z}(x,y)=\frac{\omega_{\rm B}\,\vert z\vert^{-2m_z}\exp\left(-\xi^2\right)}{4\pi\left[2 \, _0F_1\left(;-m_z+1;\vert\tau\vert^2\right)-1\right]}\times\\
		\nonumber&\quad\times\left[ \frac{1}{(-m_z)!}+\left\vert\sum_{n=1}^{\infty}\frac{\sqrt{(-m_z)!}(-\tau)^n}{(n-m_z)!}L_{n}^{-m_z}(\xi^2)\right\vert^2+\left\vert\sum_{n=1}^{\infty}\frac{\sqrt{(-m_z)!}(-\tau)^n}{(n-m_z)!}\frac{z}{\sqrt{n}}L_{n-1}^{-m_z+1}(\xi^2)\right\vert^2\right.\\
		&\quad\left.+2\Re\left[\sum_{n=1}^{\infty}\frac{(-\tau)^n}{(n-m_z)!}L_{n}^{-m_z}(\xi^2)\right]\right], \label{107a}\\
		\langle H\rangle_\tau&=\frac{2\sqrt{\omega_{\rm B}}\hbar v'_{\rm F}}{2\,_0F_1\left(;-m_z+1;\vert\tau\vert^2\right)-1}\sum_{n=0}^{\infty}\frac{(-m_z)!\,\vert\tau\vert^{2n}}{n!(n-m_z)!}\sqrt{n}, \quad {\rm for}\,m_z\leq0. \label{107c}
		\end{align}
	
\end{subequations}
\normalsize

Figures \ref{fig:rho1P_z} and \ref{fig:rho1N_z} show that the probability density of the {\it su}(1,1) coherent states (with both positive and negative $z$-component of the total angular momentum operator) always stays centered at the origin for any value of the eigenvalue $\tau$, in contrast to the bidimensional coherent states for which their eigenvalues $\alpha$ and $\beta$ determine the location of the maximum probability density on the $xy$-plane. However, for a given magnetic field strength and $\zeta$-value, both $\tau$ and $m_z$ (or $j$) modify the probability density: as the parameters $\vert\tau\vert$ and $\vert j\vert$ increase, the maximum probability moves away radially from the origin following the criterion given in Eq.~(\ref{eccentricity}). Besides, if the magnetic field strength increases, the maximum probability also does while the distance respect to origin decreases following the elliptical shape adopted, which is according to the semi-classical behavior of charged particle in a magnetic field.

On the other hand, as the $z$-component of the total angular momentum $\vert j\vert$ increases, for coherent states with $j<0$ the corresponding mean energy $\langle H\rangle_\tau$ takes values becoming smaller; meanwhile, for the states with $j>0$ the contrary effect occurs~(see Fig.~\ref{fig:H_z}).

\section{Final remarks}\label{sec5}
In this work, in order to analyze the bidimensional effects of the anisotropy in 2D Dirac materials on the dynamics of their charge carriers, we have obtained bidimensional and {\it su}$(1,1)$ coherent states through a symmetric gauge vector potential for the interaction between such particles and a homogeneous orthogonal magnetic field. Such coherent states have been obtained as eigenstates of generalized annihilation operators that satisfy either the Heisenberg-Weyl or {\it su}(1,1) algebras.

For the vector potential in Eq.~(\ref{1}), energy spectrum of the anisotropic Dirac Hamiltonian $H$ has an infinite degeneracy due to the rotational symmetry, $[\mathcal{H}^\pm,L_z]=0$, while its solutions have axial-like symmetry (Eq.~(\ref{37})). Moreover, we can identify two set of scalar ladder operators that satisfy two independent copies of the Heisenberg-Weyl algebra and with which, it is possible to define a set of generalized annihilation operators $\mathbb{A}^{-}$, $\mathbb{B}^{-}$ and $\mathbb{K}^{-}$ (Eqs.~(\ref{19a}, \ref{19b} and \ref{24}). By employing these annihilation operators, at least two different kinds of coherent states $\Psi_{\alpha,\beta}(x,y)$ and $\Phi_{\tau}^{m_z}(x,y)$ are obtained, each one with different characteristics.

The first family of coherent states, identified as 2D-CS $\Psi_{\alpha,\beta}(x,y)$, exhibits a stable shape for any value of the eigenvalues $\alpha$ and $\beta$ (Fig.~\ref{fig:rho_alphabeta1}), which determine the location of the maximum probability peak. In addition, Figure \ref{fig:rho_alphabeta} shows the anisotropy effect on these states: as $\zeta$-parameter grows, {\it i.e.}, we go from a Dirac cone aligned to the $p_{x}$-direction to one aligned to the $p_{y}$-direction, the probability density is aligned from being parallel to the $y$-axis to being parallel to $x$-axis, acquiring an elliptical or circular projection over the $xy$-plane. When $\zeta=1$, one recovers the results in Ref.~\cite{dnn19}, so we can consider those ones as a particular case of the coherent states here presented. Likewise, the second family or {\it su}(1,1)-CS $\Phi_{\tau}^{m_z}(x,y)$ possesses a probability distribution whose shape is basically an elliptical ring around the origin of coordinates (Figs. \ref{fig:rho1P_z} and \ref{fig:rho1N_z}), and whose length of the semi-major axis depends on the value of the $\zeta$-parameter. Once again, changing the anisotropy direction, charge carriers are {\it confined} to move in one or another direction, that along with a growing magnetic field strength, the probability to find such particles in a particular region on the $xy$-plane also increases.

In addition, the behavior of the mean energy values (Figs. \ref{fig:H_alpha} and \ref{fig:H_z}) suggests the possibility of using both families of coherent states in graphene in semi-classical treatments. Besides, in the limit $\alpha\rightarrow0$ and $\tau\rightarrow0$, the $\langle H\rangle_\alpha$ and $\langle H\rangle_\tau$ functions behave differently, as we can see in Figs. \ref{fig:H_alpha} and \ref{fig:H_z}. This is due to that the minimum energy states that contribute to both superpositions in Eqs.~(\ref{38}) and (\ref{105}) are those ones with $E_0=0$, while that for the linear combination in Eq.~(\ref{118}) are excited states with $E_{m_z}=\hbar\,v'_{\rm F}\sqrt{m_z\omega_{\rm B}}$.


Furthermore, it is worth to remark that in the case in which the present problem is addressed in the phase space representation, such a space is fourth-dimensional $(x,p_{x},y,p_{y})\in\mathbb{R}^{2}\times\mathbb{R}^{2}$, so that an appropriate description of the system dynamics is not trivial to obtain. The problem of a topologically cylindrical phase space $(\theta,p)\in S^{1}\times\mathbb{R}$ for a simple rotator around a fixed axis, where $S^{1}$ represents the unit circle, the position is given by an angle $\theta$ and the angular momentum by a real number $p$, has been considered in previous works~\cite{bz77,mukunda79,mmzcs05,kastrup06,rsk11,pbt14}.
	
Finally, we would like to empathize that since 2D-CS and {\it su}$(1,1)$-CS obtained here describe the same physical problem, they share information that expresses in a different way. For instance, the corresponding probability densities are deformed, showing an elliptically {\it squeezed} distribution for the former and elliptical ring for the latter, which is according with the $\zeta$-parameter that depends on the anisotropy in 2D Dirac materials. Likewise, the main difference between them is that the bidimensional coherent states allow to describe the probability of finding an electron in a small section of a closed trajectory but knowing nothing about the $z$-component of the total angular momentum, while with the {\it su}$(1,1)$ coherent states one can find the particle along the whole elliptical curve with equal probability but having meaningful information about the $z$-component of the total angular momentum. Although we have chosen a particular gauge to address the problem, the issue of the gauge dependence of the states of charged particles in background magnetic fields, and henceforth of the corresponding CS is still an open problem. First discussion on this topic has very recently been addressed in  Ref.~\cite{davighi19}. In turn, the gauge dependence of the corresponding (generalized) annihilation operators and thus of the CS is currently under consideration by our group. Results will be reported elsewhere.

%
%

%
%

%

\section*{Acknowledgments}
YCS acknowledges support from CIC-UMSNH under grant 4624705. AR acknowledges support from Consejo Nacional de Ciencia y Tecnolog\'{\i}a (M\'exico) under grant 256494. EDB acknowledges IFM-UMSNH for its warm hospitality as well as LM Nieto and J Negro for valuable discussions that motived the mathematical aspect of this work, and FJ Turrubiates for his important comments for improving it.

\appendix
\numberwithin{equation}{section}

\section{Algebraic structure and eigenstates}\label{appA}
In order to obtain the algebraic relations associated to this system, we make the following coordinate transformation
\begin{equation}
x=\zeta^{1/2}\rho\cos(\theta), \quad y=\zeta^{-1/2}\rho\sin(\theta),
\end{equation}
that corresponds to the ellipse equation
\begin{equation}
\frac{x^2}{\zeta \rho^2}+\frac{y^2}{\zeta^{-1}\rho^2}=1.
\end{equation}
Thus, by defining the dimensionless variable $\xi$ as
\begin{equation}\label{9a}
\xi=\frac{\sqrt{\omega_{\rm B}}}{2}\rho,
\end{equation}
the corresponding Hamiltonian operators in Eq.~(\ref{7}) \cite{f28,d31,dknn17} 
can be factorized in terms of two sets of differential operators, namely \cite{dknn17,df96,kka12},
\begin{equation}\label{17}
\mathcal{H}^+=A^{+}A^{-}=B^{+}B^{-}+L_z, \quad \mathcal{H}^-=\mathcal{H}^++1,
\end{equation}
where
\begin{subequations}\label{13}
	\begin{align}
	A^{-}&=\frac{\exp(-i\theta)}{2}\left(\partial_\xi-\frac{i\partial_\theta}{\xi}+\xi\right), \quad A^{+}=(A^{-})^\dagger, \label{20b} \\ 
	B^{-}&=\frac{\exp(i\theta)}{2}\left(\partial_\xi+\frac{i\partial_\theta}{\xi}+\xi\right), \qquad B^{+}=(B^{-})^\dagger, \label{21b} \\ 
	L_z&=-i\partial_{\theta}=N-M, \label{21c}
	\end{align}
\end{subequations}
being $L_z=\left(\zeta^{-1}xp_y-\zeta yp_x\right)/\hbar$ the $z$-component of an angular momentum-like operator and $N=A^{+}A^{-}$ and $M=B^{+}B^{-}$ are number operators. The above operators satisfy the following commutation relations
\begin{subequations}\label{22}
	\begin{align}
	[A^{-},A^{+}]&=\mathbf{1}, & [B^{-},B^{+}]&=\mathbf{1}, \label{13a} \\
	[A^{\pm},B^{\pm}]&=\mathbf{0}, & [A^{\pm},B^{\pm}]&=\mathbf{0}, \label{13b} \\
	[L_z,A^{\pm}]&=\pm A^{\pm}, & [L_z,B^{\pm}]&=\mp B^{\pm}. \label{13c}
	\end{align}
\end{subequations}
These relations imply that each set of ladder operators $A^{\pm}$ and $B^{\pm}$ are generators of the Heisenberg-Weyl algebra and are also independent of each other.

Eq.~(\ref{13c}) implies that the operators $A^{+}$ and $A^{-}$, upon acting on an eigenstate of $L_z$, increase or decrease, respectively, the eigenvalue of $L_z$ in an unity; meanwhile, the operators $B^{\pm}$ have the contrary effect.

\begin{figure}[!ht]
	\begin{center}
		\includegraphics[width=.45\columnwidth]{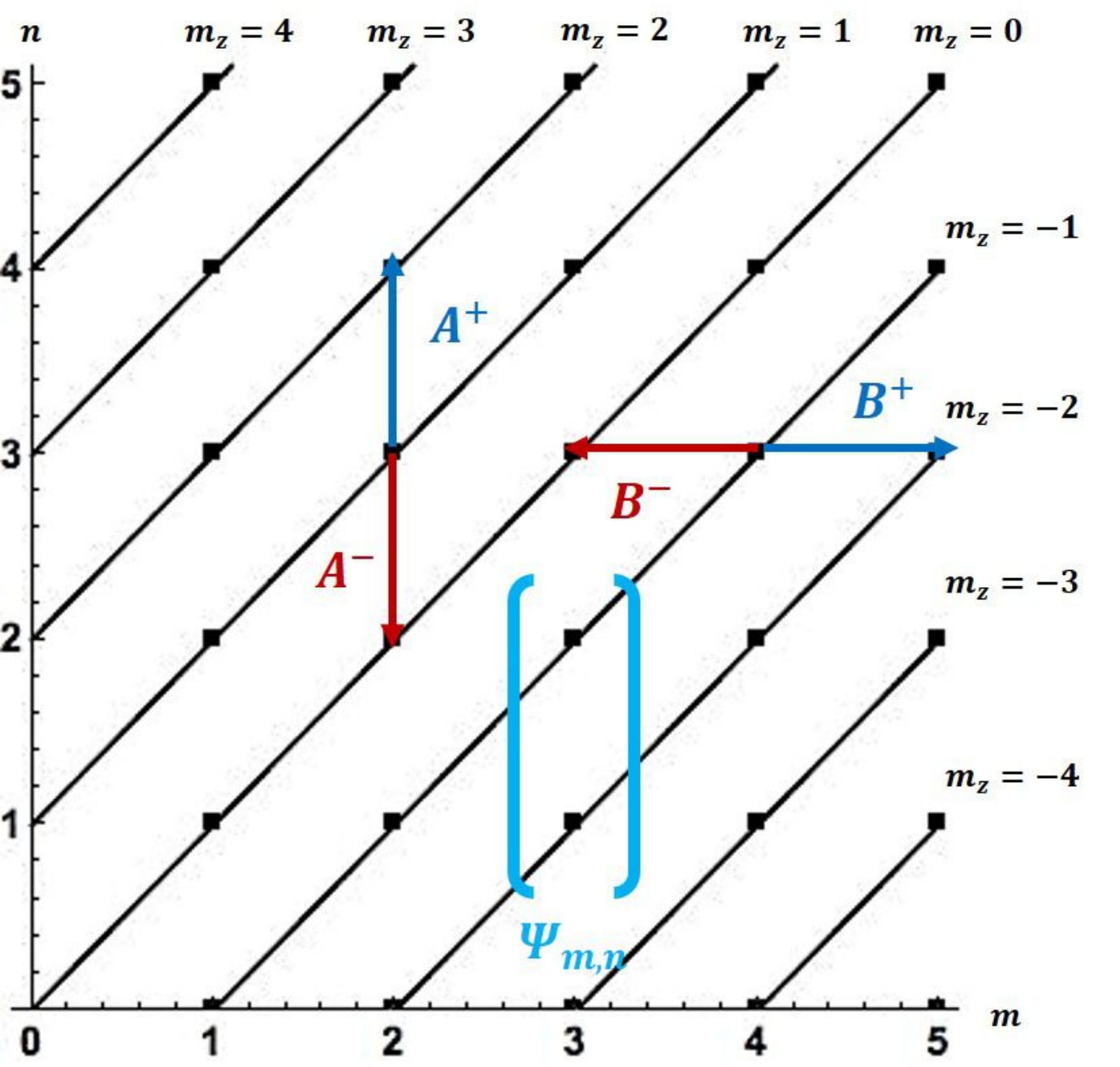}
	\end{center}
	\caption{\label{fig:diagram}Space of scalar states $\psi_{m,n}$ is represented univocally by coordinates $(m,n)$. Inclined lines connect states with the same angular momentum $m_z=n-m$. The plane is divided in two sectors according to $m_z\geq0$ (upper sector) or $m_z\leq0$ (lower sector).}
\end{figure}

In addition, the action of the operators $A^{\pm}$ and $B^{\pm}$ on the states $\psi_{m,n}$ is (see Fig.~\ref{fig:diagram})
\begin{subequations}\label{23}
	\begin{align}
	A^{-}\psi_{m,n}=\sqrt{n}\,\psi_{m,n-1}, &\quad A^{+}\psi_{m,n}=\sqrt{n+1}\,\psi_{m,n+1}, \label{23a}\\
	B^{-}\psi_{m,n}=\sqrt{m}\,\psi_{m-1,n}, &\quad B^{+}\psi_{m,n}=\sqrt{m+1}\,\psi_{m+1,n}. \label{23b}
	\end{align}
\end{subequations}

Furthermore, because of the choice of gauge, rotational symmetry $[\mathcal{H}^{\pm},L_{z}]=0$ is present and the scalar solutions $\psi_{m,n}$ in Eq.~(\ref{34}) have axial symmetry and infinite degeneracy. In the sense of dynamical systems integrability, the addressed problem is said to be integrable due to the symmetry operator $L_{z}$, leading to the $z$-component of the angular momentum-like operator being a conserved quantity.

\section{Classification of eigenstates}\label{class}
We can label as $\Psi_{m,n}^{+}(x,y)$ the spinor states whose two scalar components have positive $z$-component of the angular momentum $(m_z\geq0)$, and as $\Psi_{m,n}^{-}(x,y)$ those ones whose two scalar components have negative $z$-component angular momentum $(m_z\leq0)$, {\it i.e.},
\begin{subequations}\label{37}
	\begin{align}
	\Psi_{m,n}^{+}(x,y)&=\frac{1}{\sqrt{2}}\left(\begin{array}{c}
	\psi_{m,n-1}^{+}(x,y)\\
	i\psi_{m,n}^{+}(x,y)
	\end{array}\right), \\
	\Psi_{m,n}^{-}(x,y)&=\frac{1}{\sqrt{2^{(1-\delta_{0n})}}}\left(\begin{array}{c}
	(1-\delta_{0n})\psi_{m,n-1}^{-}(x,y)\\
	i\psi_{m,n}^{-}(x,y)
	\end{array}\right),
	\end{align}
\end{subequations}
where $\psi^+_{m,n}(x,y)\equiv\psi_{m,n}^+(\rho,\theta)$  ($\psi^-_{m,n}(x,y)\equiv\psi_{m,n}^-(\rho,\theta)$) identifies the states that belong to the upper (lower) sector in Fig. \ref{fig:diagram}, and $\delta_{mn}$ denotes the Kronecker delta.

Hence, the corresponding Hilbert space spanned by the spinors $\vert\Psi_{m,n}\rangle$ in Dirac notation is given by
	\begin{equation}
\mathcal{H}={\rm span}\{\vert\Psi_{m,n}\rangle\vert\, n, m\in\mathbb{Z}^{+}\cup\{0\}\},
	\end{equation}
	such that the identity operator $\mathbf{1}$ on $\mathcal{H}$ writes as
	\begin{equation}
	\mathbf{1}=\sum_{m=0}^{\infty}\vert \Psi_{m,0}\rangle\langle \Psi_{m,0}\vert +\sum_{n=1}^{\infty}\sum_{m=0}^{\infty}\vert \Psi_{m,n}\rangle\langle \Psi_{m,n}\vert.
	\end{equation}
	
	Besides, the space $\mathcal{H}$ can be also decomposed in two subspaces according to the quantum number $m_{z}=n-m$ as follows~\cite{aremua15}:
	\begin{equation}
	\mathcal{H}=\left(\bigoplus_{m'_{z}\leq0}\mathcal{H}_{m'_{z}}\right)\bigoplus\left(\bigoplus_{m_{z}=1}^{\infty}\mathcal{H}_{m_{z}}\right),
	\end{equation}
	in order to guaranty $j<0$ and $j>0$, respectively. Likewise, the identity operator on each subspace turns out to be:
	\begin{equation}\label{identity}
	\mathbf{1}=\left(\bigoplus_{m'_{z}\leq0}\mathbf{1}_{m'_z}\right)\bigoplus\left(\bigoplus_{m_{z}=1}^{\infty}\mathbf{1}_{m_z}\right),
	\end{equation}
	where
\begin{subequations}
	\begin{align}
	\mathbf{1}_{m'_z}&=\vert\Phi_{m'_z,0}\rangle\langle\Phi_{m'_z,0}\vert+\sum_{n=1}^{\infty}\vert \Phi_{m'_z,n}\rangle\langle\Phi_{m'_z,n}\vert, \\
	\mathbf{1}_{m_z}&=\sum_{n=m_z}^{\infty}\vert\Phi_{m_z,n}\rangle\langle\Phi_{m_z,n}\vert.
	\end{align}
\end{subequations}

\section{Algebra generated by matrix operators}\label{algebra}
The operators $\mathbb{A}^{-}$ and $\mathbb{B}^{-}$ satisfy two independent copies of the Heisenberg-Weyl algebra:
\begin{subequations}
	\begin{align}
	[\mathbb{A}^{-},\mathbb{A}^{+}]&=\mathbb{I}, \quad \mathbb{A}^{+}=(\mathbb{A}^{-})^\dagger, \label{45a}\\
	[\mathbb{B}^{-},\mathbb{B}^{+}]&=\mathbb{I},  \quad \mathbb{B}^{+}=(\mathbb{B}^{-})^\dagger, \label{45b}\\
	[\mathbb{A}^{\pm},\mathbb{B}^{\pm}]&=[\mathbb{A}^{\pm},\mathbb{B}^{\mp}]=\mathbf{0}. \label{45c}
	\end{align}
\end{subequations}

Likewise, considering $\mathbb{K}^{+}=(\mathbb{K}^{-})^\dagger$, we get
\begin{subequations}
	\begin{align}
	[\mathbb{K}^{-},\mathbb{K}^{+}]=2\mathbb{K}_{0}&=\left[\begin{array}{c c}
	N+M+2 & 0 \\
	0 & N+M+1
	\end{array}\right], \\
	[\mathbb{K}_{0},\mathbb{K}^{\pm}]&=\pm\mathbb{K}^{\pm}.
	\end{align}
\end{subequations}
which allow us to identify an {\it su$(1,1)$} algebra.

Similar to what happens with scalar counterparts, these matrix operators are closely related to conserved observables that are considered significant. For instance, we take into account the operators $\mathbb{A}^{\pm}$ and $\mathbb{B}^{\pm}$ to obtain quantum states with a well-defined level energy, while to construct states with a well-defined $z$-component of the angular momentum-like operator $\mathbb{J}_{z}$, we define the operators $\mathbb{K}^{\pm}=\mathbb{A}^{\pm}\mathbb{B}^{\pm}$, which turn out to be a boson realization of $su(1,1)$ algebra~\cite{doebner89,LoLiu93,Wunsche99,rosas16}, and that in turn is a particular case of the Schwinger boson representation~\cite{gazeau10}. In this sense, it is worth to mention that a realization with matrix operators that are generator of $su(2)$ algebra could be performed~\cite{moran2019} in the context of this work, in order to explore different conserved observable.

\section{Obtaining the 2D-CS}\label{2DCS}
In order to obtain the simplest form for bidimensional coherent states, we  proceed to build the states $\Psi_{\beta}^n(x,y)$ for the case with $\eta=0$, so that $\mathbb{B}^{-}=B^{-}\otimes\mathbb{I}$. From the corresponding eigenvalue equation~(\ref{29}), the coherent states $\Psi_{\beta}^n(x,y)$ turn out to be
\begin{equation}\label{30}
\Psi_\beta^n(x,y)=
\frac{1}{\sqrt{2^{(1-\delta_{0n})}}}\left(\begin{array}{c}
(1-\delta_{0n})\psi_\beta^{n-1}(x,y) \\
i\psi_\beta^{n}(x,y)
\end{array}\right),
\end{equation}
where $\psi_\beta^n(x,y)$ and $\psi_\beta^{n-1}(x,y)$ are identified as scalar eigenstates of the operator $B^{-}$ for each $n$ index, {\it i.e.}, these states are stationary  with a well-defined energy $E_n$.

Now, one can notice that each of these coherent states satisfies one of the following equation systems,
\begin{equation}\label{31}
\left\{\begin{aligned}
B^{-}\psi_\beta^n&=\beta\psi_\beta^n,\\
A^{+}A^{-}\psi_\beta^n&=n\,\psi_\beta^n,
\end{aligned}
\right. \qquad
\left\{\begin{aligned}
B^{-}\psi_\beta^{n-1}&=\beta\psi_\beta^{n-1},\\
A^{-}A^{+}\psi_\beta^{n-1}&=n\,\psi_\beta^{n-1}.
\end{aligned}
\right.
\end{equation}

Thus, after solving the systems of equations~(\ref{31}), the normalized coherent states $\Psi_\beta^n(x,y)$, $n=0,1,2,\dots,$ are
\begin{equation}
\Psi_\beta^n(x,y)=\frac{1}{\sqrt{2^{(1-\delta_{0n})}\pi\,n!}}\exp\left(\left[\beta-\frac{z}{2}\right]z^\ast-\frac{\vert\beta\vert^2}{2}\right)
\left(\begin{array}{c}
\sqrt{n}(z-\beta)^{n-1}\\
i(z-\beta)^n
\end{array}\right),\label{33}
\end{equation}
where the complex parameter $z$ is defined by
\begin{equation}\label{32}
z=\frac{\sqrt{\omega_{\rm B}}}{2}\left(\zeta^{-1/2}x+i\zeta^{1/2}y\right)=\frac{\sqrt{\omega_{\rm B}}}{2}\rho\exp(i\theta).
\end{equation}

Finally, by taking the sum over the $n$ index, one gets the expression in~(\ref{38}).


\section{Completeness relation}\label{completeness}
We investigate the non-orthogonality property and the resolution of unity of both the bidimensional and {\it su}$(1,1)$ coherent states, setting $\delta=0$ by simplicity.

First, the coherent states in Eq.~(\ref{38}) can be also rewritten, in Dirac notation, as
\begin{equation}
\vert\Psi_{\alpha,\beta}\rangle=\frac{\exp\left(-\frac{\vert\beta\vert^2}{2}\right)}{\sqrt{2\exp(\vert\alpha\vert^2)-1}}\left[\sum_{m=0}^{\infty}\frac{\beta^m}{\sqrt{m!}}\vert\Psi_{m,0}\rangle+\sum_{n=1}^{\infty}\sum_{m=0}^{\infty}\frac{\sqrt{2}\alpha^n\beta^m}{\sqrt{n!m!}}\vert\Psi_{m,n}\rangle\right].
\end{equation}

Taking the inner product between two 2D-CS with different eigenvalues, we have that:
\begin{equation}
\vert\langle\Psi_{\alpha',\beta'}\vert\Psi_{\alpha,\beta}\rangle\vert=\Bigg\vert\exp\left(-\frac{\vert\beta'\vert^2}{2}\right)\exp\left(-\frac{\vert\beta\vert^2}{2}\right)\frac{\left(2\exp\left(\alpha'^\ast\alpha\right)-1\right)\exp\left(\beta'^\ast\beta\right)}{\sqrt{2\exp(\vert\alpha'\vert^2)-1}\sqrt{2\exp(\vert\alpha\vert^2)-1}}\Bigg\vert,
\end{equation}
which implies the non-orthogonality property of the 2D-CS. For $\alpha=\alpha'$ and $\beta=\beta'$, one recovers the normalization relationship $\langle\Psi_{\alpha,\beta}\vert\Psi_{\alpha,\beta}\rangle=1$.

Now, choosing the measure
\begin{align}\label{E3} d\mu(\alpha,\beta)&=\frac{2\exp(\vert\alpha\vert^2)-1}{2\pi^2}\vert\alpha\vert\vert\beta\vert e^{-\vert\alpha\vert^2}d\vert\alpha\vert d\vert\beta\vert d\theta d\varphi,
\end{align}
it follows that
\small

	\begin{align}\label{E4}
	\nonumber &\sum_{m=0}^{\infty}\frac{\vert \Psi_{m,0}\rangle\langle \Psi_{m,0}\vert }{2}+\int_{\mathbb{C}_{\alpha}}\int_{\mathbb{C}_{\beta}}\vert \Psi_{\alpha,\beta}\rangle\langle \Psi_{\alpha,\beta}\vert d\mu(\alpha,\beta) =\sum_{m=0}^{\infty}\frac{\vert \Psi_{m,0}\rangle\langle\Psi_{m,0}\vert }{2}+\int_{\mathbb{C}_{\alpha}}\int_{\mathbb{C}_{\beta}}\frac{d\mu(\alpha,\beta)\exp\left(-\vert\beta\vert^2\right)}{2\exp(\vert\alpha\vert^2)-1} \\
	\nonumber &\quad\times\left[\sum_{m=0}^{\infty}\frac{\beta^m}{\sqrt{m!}}\vert \Psi_{m,0}\rangle+\sum_{n=1}^\infty\sum_{m=0}^{\infty}\frac{\sqrt{2}\alpha^{n}\beta^m}{\sqrt{n!m!}} \vert \Psi_{m,n}\rangle\right]\left[\sum_{m'=0}^{\infty}\frac{\beta^{\ast m'}}{\sqrt{m'!}}\langle \Psi_{m',0}\vert +\sum_{n'=1}^\infty\sum_{m'=0}^{\infty} \frac{\sqrt{2}\alpha^{\ast n'}\beta^{\ast m'}}{\sqrt{n'!m'!}} \langle \Psi_{m',n'}\vert \right] \\
	\nonumber &=\sum_{m=0}^{\infty}\frac{\vert \Psi_{m,0}\rangle\langle \Psi_{m,0}\vert }{2}+\frac{1}{2}\sum_{m=0}^{\infty}\int_{0}^{\infty}\frac{t^{m}e^{-t}}{m!}dt\int_{0}^{\infty}\left[ \vert \Psi_{m,0}\rangle\langle \Psi_{m,0}\vert +2\sum_{n=1}^{\infty} \frac{\vert \Psi_{m,n}\rangle\langle \Psi_{m,n}\vert }{n!} s^{n}\right]e^{-s}ds \\
	\nonumber &=\sum_{m=0}^{\infty}\frac{\vert \Psi_{m,0}\rangle\langle \Psi_{m,0}\vert }{2}+\sum_{m=0}^{\infty}\frac{\vert \Psi_{m,0}\rangle\langle \Psi_{m,0}\vert }{2}+\sum_{n=1}^{\infty}\sum_{m=0}^{\infty}\vert \Psi_{m,n}\rangle\langle \Psi_{m,n}\vert \\
	&=\sum_{m=0}^{\infty}\vert \Psi_{m,0}\rangle\langle \Psi_{m,0}\vert +\sum_{n=1}^{\infty}\sum_{m=0}^{\infty}\vert \Psi_{m,n}\rangle\langle \Psi_{m,n}\vert =\mathbf{1},
	\end{align}

\normalsize
where $t=\vert\beta\vert^2$, $s=\vert\alpha\vert^2$ and we use the following results
\begin{subequations}
	\begin{align}
	&\int_{0}^{2\pi}\exp\left(i(p-q)\chi\right)d\chi=2\pi\delta_{pq}, \\
	&\int_{0}^{\infty}x^ke^{-x}dx=\Gamma(k+1)=k!.
	\end{align}
\end{subequations}

On the other hand, the {\it su}$(1,1)$-CS with $j>0$ in Eq.~(\ref{105}) are expressed in Dirac notation as
\small
\begin{equation}\label{56}
\vert\Phi_{\tau}^{m_z}\rangle=\frac{1}{\sqrt{_0F_1\left(;m_z+1;\vert\tau\vert^2\right)}}\sum_{n=m_z}^{\infty}\frac{\sqrt{m_z!}\,\tau^{n-m_z}}{\sqrt{n!(n-m_z)!}}\vert\Phi_{m_z,n}\rangle.
\end{equation}
\normalsize

The inner product of two coherent states with different eigenvalues turns out to be:
\small
\begin{equation}
\vert\langle\Phi_{\tau'}^{m_z}\vert\Phi_\tau^{m_z}\rangle\vert=\left\vert\frac{_0F_1\left(;m_z+1;\tau'^{\ast}\tau\right)}{\sqrt{_0F_1\left(;m_z+1;\vert\tau'\vert^2\right)\,_0F_1\left(;m_z+1;\vert\tau\vert^2\right)}}\right\vert.
\end{equation}
\normalsize
This expression indicates the non-orthogonality property of the {\it su}$(1,1)$-CS. When $\tau=\tau'$, we obtain again the normalization condition $\langle\Phi_{\tau}^{m_z}\vert\Phi_{\tau}^{m_z}\rangle=1$.

Now, considering the measure
\begin{equation}\label{E8}
d\mu(\tau)=\frac{_0F_1\left(;m_z+1;\vert\tau\vert^2\right)}{\pi}f(\vert\tau\vert)\vert\tau\vert d\vert\tau\vert d\theta,
\end{equation}
we get for a $m_z$ given:
\footnotesize
	\begin{align}\label{E9}
	\nonumber\int_{\mathbb{C}}\vert\Phi_{\tau}^{m_z}\rangle\langle\Phi_{\tau}^{m_z}\vert d\mu(\tau)&=\int_{\mathbb{C}}\frac{d\mu(\tau)}{_0F_1\left(;m_z+1;\vert\tau\vert^2\right)}\left[\sum_{n=m_z}^{\infty}\frac{\sqrt{m_z!}\,\tau^{n-m_z}}{\sqrt{n!(n-m_z)!}}\vert\Phi_{m_z,n}\rangle\right]\left[\sum_{n'=m_z}^{\infty}\frac{\sqrt{m_z!}\,\tau^{\ast n'-m_z}}{\sqrt{n'!(n'-m_z)!}}\langle\Phi_{m_z,n'}\vert\right] \\
	\nonumber&
	=\sum_{n=m_z}^{\infty}\frac{m_z!\vert\Phi_{m_z,n}\rangle\langle\Phi_{m_z,n}\vert}{n!(n-m_z)!}\int_{0}^{\infty}t^{n-m_z}f(t)dt \\
	&
	=\sum_{n=m_z}^{\infty}\vert\Phi_{m_z,n}\rangle\langle\Phi_{m_z,n}\vert=\mathbf{1}_{m_z},
	\end{align}
\normalsize
where $t=\vert\tau\vert^2$ and $f(t)$ satisfies the moment problem:
	\begin{equation}
	\int_{0}^{\infty}t^{s-m_z}f(t)dt=\int_{0}^{\infty}t^{s-1}F(t)dt=\frac{\Gamma(s+1)\Gamma(s-m_z+1)}{\Gamma(m_z+1)}\equiv\mathcal{F}(s),
	\end{equation}
	where $F(t)=t^{1-m_z}f(t)$ and $\mathcal{F}(s)\equiv\mathcal{M}_{t}[F(t)](s)$ denotes the Mellin transform.
	
By employing the inverse Mellin transform, we find~\cite{fhn94}:
	\begin{equation}
	F(t)=\mathcal{M}_{s}^{-1}\left[\mathcal{F}(s)\right](t)=\frac{1}{2\pi i}\int_{\gamma-i\infty}^{\gamma+i\infty}t^{-s}\mathcal{F}(s)ds=\frac{2 t^{1-m_{z}/2} K_{-m_{z}}\left(2 \sqrt{t}\right)}{\Gamma (m_{z}+1)},
	\end{equation}
	where $K_{n}(z)$ denotes the modified Bessel function of the second kind. Since $K_{-m}(x)=K_{m}(x)$, the function $f(\vert\tau\vert)$ in the measure (\ref{E8}) turns out to be:
	\begin{equation}
	f(\vert\tau\vert)=\frac{2\vert\tau\vert^{m_{z}}K_{m_{z}}\left(2 \vert\tau\vert\right)}{\Gamma (m_{z}+1)}.
	\end{equation}

Finally, the {\it su}$(1,1)$-CS with $j<0$ in Eq.~(\ref{118}) read now as
\begin{equation}\label{61}
\vert\Phi_{\tau}^{m_z}\rangle=\frac{1}{\sqrt{2 \, _0F_1\left(;-m_z+1;\vert \tau\vert^2\right)-1}}\left[\vert\Phi_{m_z,0}\rangle+\sum_{n=1}^{\infty}\frac{\sqrt{2\,(-m_z)!}\tau^n}{\sqrt{n!(n-m_z)!}}\vert\Phi_{m_z,n}\rangle\right],
\end{equation}
whose inner product between CS with different eigenvalues is
\small
\begin{equation}
\vert\langle\Phi_{\tau'}^{m_z}\vert\Phi_\tau^{m_z}\rangle\vert=\Bigg\vert\frac{2\,_0F_1\left(;-m_z+1;\tau'^{\ast}\tau\right)-1}{\sqrt{2\,_0F_1\left(;-m_z+1;\vert\tau'\vert^2\right)-1}\sqrt{2\,_0F_1\left(;-m_z+1;\vert\tau\vert^2\right)-1}}\Bigg\vert.
\end{equation}
\normalsize
Once again, this expression indicates the non-orthogonality property of the corresponding {\it su}$(1,1)$-CS and the normalization condition $\langle\Phi_{\tau}^{m_z}\vert\Phi_{\tau}^{m_z}\rangle=1$ is obtained for $\tau=\tau'$.

Assuming the following measure
\begin{equation}\label{E16}
d\mu(\tau)=\frac{2\,_0F_1\left(;-m_z+1;\vert\tau\vert^2\right)-1}{2\pi}g(\vert\tau\vert)\vert\tau\vert d\vert\tau\vert d\theta,
\end{equation}
we have that

	\begin{align}\label{E14}
	\nonumber&\frac{\vert\Phi_{m_z,0}\rangle\langle\Phi_{m_z,0}\vert }{2}+\int_{\mathbb{C}}\vert \Phi_{\tau}^{m_z}\rangle\langle \Phi_{\tau}^{m_z}\vert d\mu(\tau)=\frac{\vert\Phi_{m_z,0}\rangle\langle\Phi_{m_z,0}\vert }{2}+\int_{\mathbb{C}}\frac{d\mu(\alpha)}{2\,_0F_1\left(;-m_z+1;\vert\tau\vert^2\right)-1} \\
	\nonumber &\nonumber\quad\times\left[\vert\Phi_{m_z,0}\rangle+\sum_{n=1}^{\infty}\frac{\sqrt{2\,(-m_z)!}\tau^n}{\sqrt{n!(n-m_z)!}}\vert\Phi_{m_z,n}\rangle\right]\left[\langle\Phi_{m_z,0}\vert+\sum_{n'=1}^{\infty}\frac{\sqrt{2\,(-m_z)!}\tau^{\ast n'}}{\sqrt{n'!(n'-m_z)!}}\langle\Phi_{m_z,n'}\vert\right] \\
	\nonumber &=\frac{\vert\Phi_{m_z,0}\rangle\langle\Phi_{m_z,0}\vert }{2}+\int_{0}^{\infty}\left[\frac{\vert\Phi_{m_z,0}\rangle\langle\Phi_{m_z,0}\vert}{2}+\sum_{n=1}^{\infty}\frac{(-m_z)!\,\vert \Phi_{m_z,n}\rangle\langle \Phi_{m_z,n}\vert }{n'!(n-m_z)!}t^{n}\right]g(t)dt \\
	\nonumber &=\frac{\vert\Phi_{m_z,0}\rangle\langle\Phi_{m_z,0}\vert }{2}+\frac{\vert\Phi_{m_z,0}\rangle\langle\Phi_{m_z,0}\vert }{2}\int_{0}^{\infty}g(t)dt+\sum_{n=1}^{\infty}\frac{(-m_z)!\vert \Phi_{m_z,n}\rangle\langle \Phi_{m_z,n}\vert}{n!(n-m_z)!}\int_{0}^{\infty}t^{n}g(t)dt \\
	&
	=\vert\Phi_{m_z,0}\rangle\langle\Phi_{m_z,0}\vert+\sum_{n=1}^{\infty}\vert \Phi_{m_z,n}\rangle\langle\Phi_{m_z,n}\vert=\mathbf{1}_{m_z},
	\end{align}

where $t=\vert\tau\vert^2$ and $g(t)$ satisfies the corresponding moment problem:
	\begin{equation}
	\int_{0}^{\infty}t^{s}g(t)dt=\int_{0}^{\infty}t^{s-1}G(t)dt=\frac{\Gamma(s+1)\Gamma(s-m_z+1)}{\Gamma(-m_z+1)}\equiv\mathcal{G}(s),
	\end{equation}
	where $G(t)=t\,g(t)$ and $\mathcal{G}(s)\equiv\mathcal{M}_{t}[G(t)](s)$. The inverse Mellin transform allows us to find
	\begin{equation}
	G(t)=\mathcal{M}_{s}^{-1}\left[\mathcal{G}(s)\right](t)=\frac{1}{2\pi i}\int_{\gamma-i\infty}^{\gamma+i\infty}t^{-s}\mathcal{G}(s)ds=\frac{2 t^{1-m_{z}/2} K_{-m_{z}}\left(2 \sqrt{t}\right)}{\Gamma (-m_{z}+1)}.
	\end{equation}

Therefore, the function $g(\vert\tau\vert)$ in (\ref{E16}) has the explicitly form:
\begin{equation}
g(\vert\tau\vert)=\frac{2\vert\tau\vert^{-m_{z}}K_{-m_{z}}\left(2\vert\tau\vert\right)}{\Gamma (-m_{z}+1)}.
\end{equation}

Finally, putting together the sums over the $m_{z}$ index of Eqs.~(\ref{E9}) and (\ref{E14}), and recalling that $m_{z}=n-m$, we recover the completeness relation (\ref{identity}).

\bibliographystyle{ieeetr}
\bibliography{biblio}

\end{document}